\documentclass[fleqn,subeqn,12pt]{article}
\usepackage{amsmath,amsfonts,epsfig,mathrsfs,todonotes,yfonts}
\usepackage{color}
\usepackage{cite}
\usepackage{stmaryrd}
\usepackage[bbgreekl]{mathbbol}
\numberwithin{equation}{section}

\def\+{{+\!\!\!+}} 

\newcommand{\ber}{\begin{eqnarray}}
\newcommand{\eer}{\end{eqnarray}}
\newcommand{\beq}{\begin{equation}}
\newcommand{\eeq}{\end{equation}}

\newcommand{\nn}{\nonumber}
\newcommand{\na}{\nabla}
\newcommand{\half}{{\textstyle{\frac12}}}

\newcommand{\re}[1] {(\ref{#1})}

\begin{document}

\begin{titlepage}

\begin{flushright} \small
UUITP-46/18\\
Imperial-TP-2018-CH-02\\
\end{flushright}
\smallskip
\begin{center} 
\LARGE

{\bf The Generalised Complex Geometry of    $(p,q)$ Hermitian Geometries}
 \\[30mm] 
 
\large
{\bf Chris Hull$^a$}~and~{\bf Ulf~Lindstr\"om$^{ab}$} \\[20mm]
{ \small\it
$^a$The Blackett Laboratory, Imperial College London\\
Prince Consort Road London SW7 @AZ, U.K.\smallskip

$^b$Department of  Physics and Astronomy,\\ Division of Theoretical Physics,
Uppsala University,\\
 Box 516, SE-751 20 Uppsala, Sweden\\ }

\end{center}

\vspace{10mm}
\centerline{\bfseries Abstract} 
\noindent
We define  $(p,q)$ hermitian geometry as the target space geometry of the two dimensional $(p,q)$ supersymmetric sigma model.
This includes generalised K\" ahler geometry for $(2,2)$, generalised hyperk\" ahler geometry for $(4,2)$, strong K\" ahler with torsion geometry for $(2,1)$ and
strong hyperk\" ahler with torsion geometry for $(4,1)$.
We provide a   generalised complex geometry formulation of hermitian geometry, generalising Gualtieri's formulation of the $(2,2)$ case.
Our formulation involves a chiral version of generalised complex structure and we provide explicit formulae for the map to generalised geometry.
\bigskip

\end{titlepage}

\setcounter{footnote}{0}

\tableofcontents

\section{Introduction}

Complex geometries with torsion arise in the study of supersymmetric sigma models and in generalised complex geometry.
The
bihermitean geometry that arises in  two-dimensional supersymmetric sigma models \cite{Gates:1984nk} was formulated in the framework of generalised complex geometry
\cite{Hitchin:2004ut} as
generalised K\"ahler geometry by Gualtieri in \cite{Gualtieri:2003dx}, with the relation between the sigma model geometry and generalised complex geometry
 given by the Gualtieri map.
 
 The supersymmetry algebras in two dimensions are labelled by two integers $p,q$ and different geometries arise for different values of $(p,q)$. We will refer to these here as $(p,q)$ hermitian geometries; these will be defined in section 2. 
 The models of  \cite{Gates:1984nk}  giving rise to generalised K\"ahler geometry have $(2,2)$ supersymmetry while $(2,1)$ supersymmetry gives an interesting geometry 
 \cite{Hull:1985jv, Hull:1985zy} that is
 sometimes referred to as Strong K\"ahler with Torsion (SKT).  SKT geometry was discussed within the formalism of  generalised complex geometry
in \cite{Gil1,Cavalcanti:2012fr}. Our purpose here is to find the generalised complex geometry
formulation of the  all $(p,q)$ hermitian geometries. This will require the definition of a chiral form of generalised complex structure that we will refer to as a half  generalised complex structure. We also aim to give a presentation that is closely related to the sigma model geometry, making the map to generalised complex geometry manifest.

\section{$(p,q)$ Hermitian Geometry}
\label{herman}
 The $(p,q)$ supersymmetry algebra in two dimensions has $p$ right-handed supercharges and $q$ left-handed ones \cite{Hull:1985jv}.  The general supersymmetric sigma models with $(1,1)$, $(2,2)$ and $(4,4$) supersymmetry were constructed in 
 \cite{Gates:1984nk}, the ones with $(1,0)$ and $(2,0)$ supersymmetry were constructed in \cite{Hull:1985jv}, the one with (2,1) supersymmetry was constructed in \cite{Hull:1985zy}, while the remaining cases were given in \cite{Hull:1986hn}.
The $(1,1)$ supersymmetric sigma model has a target space $({\cal M},g,H)$  which is a manifold ${\cal M}$ with a metric $g$ and a closed 3-form $H$. This can be given locally in terms of a 2-form potential $b$, $H=db$. 
The $(1,1)$ model will in fact have $(p,q)$ supersymmetry with $p,q=1,2 $ or $4 $ if it has a special geometry that we will call a $(p,q)$ {\it hermitian geometry}, which  
has $p-1$ complex structures
$J_+^a$ ($a=1,\dots p-1$) and $q-1$ complex structures
$J_-^{a'}$ ($a'=1,\dots q-1$). 
 The space $({\cal M},g,H , J_+^a, J_-^{a'})$  is a $(p,q)$ hermitian geometry if
\begin{enumerate}
\item $J_+^a$ ($a=1,\dots p-1$) and  
$J_-^{a'}$ ($a'=1,\dots q-1$) are complex structures on $ {\cal M}$.
\item The metric $g$ is hermitian with respect to all complex structures 
\ber g(J X, J Y)= g(X,Y)~.
\nonumber
\eer
\item The
 $J_\pm$ are covariantly constant 
\ber
\nonumber
\na^{(\pm)}J_{(\pm)}=0
\eer
with respect to the connections 
\ber\label{torcon}
\na^{(\pm)}:=(\na^{(0)}\pm\half g^{-1}H)
\eer
 with torsion $\pm \frac 1 2 g^{il}H_{ljk}$. Here  $\na^{(0)}$  is the  Levi-Civita connection.
\item  If $p=4$  then $J_+^1, J_+^2, J_+^3$ satisfy a quaternion algebra and if $q=4$  then $J_-^{1}, J_-^{2} , J_-^{3}  $ satisfy a quaternion algebra.
\end{enumerate}

For  each complex structure, there is a differential operator $d^c= i(\partial - \bar \partial)$ and the corresponding 2-form $\omega = gJ$ which satisfies $dd^c\omega = 0$.
The condition (3) can be replaced with the following condition:
\begin{itemize}
\item  For each $J_+$, $d^c\omega = H$ while for each $J_-$, $d^c\omega = -H$.
\end{itemize}

The $(2,2)$ geometries were called generalised K\" ahler geometries in \cite{Gualtieri:2003dx} while the $(4,2) $ gives geometries that were called generalised hyperk\" ahler geometries in \cite{bur}. 
The name strong K\"ahler with torsion (SKT) was proposed 
for $(2,1)$ (or $(1,2)$) geometry and strong hyper\"ahler with torsion was proposed for $(4,1)$ (or $(1,4)$) geometry
in \cite{Howe:1996kj}.
If $H=0$, then $p=q$ and the $(2,2)$ case gives K\" ahler geometry in which case $\omega $ is the 
K\" ahler  form,
while the $(4,4)$ case gives hyperk\" ahler geometries.

There is a rich interplay between $(p,q)$ hermitian geometry and the superspace formulation of the $(p,q)$ supersymmetric sigma model. For the  $(2,2)$ case, if the two complex structures $J_+,J_-$ commute, then the corresponding sigma model is formulated in terms of chiral and twisted chiral superfields  \cite{Gates:1984nk}.
On the other hand,  if  the commutator $[J_+,J_-]$ has trivial kernel, then  the sigma model is formulated in terms  of semichiral superfields  \cite{Buscher:1987uw}. It was shown in \cite{Lindstrom:2005zr}   that the general $(2,2)$ sigma model can be formulated in terms of chiral,  twisted chiral and semichiral superfields, giving a complete characterisation of generalised K\" ahler geometry (away from irregular points) with the geometry given locally by a scalar potential.
This will be discussed further in section 10.

\section{Generalised  Geometry}
\label{GG}
For a $d$-dimensional manifold ${\cal M}$, 
the generalised tangent space is the sum of the tangent bundle $T$ and the cotangent bundle $T^*$
\ber
\mathbb {T}:=T \oplus T^* ~.
\eer
This has a natural $O(d,d)$ invariant metric $\eta$ defined by
\ber
\eta ({X+\xi,Y+\sigma })=
\xi(Y)+\sigma(X)
\eer
where $X,Y$ are vector fields and 
  and $\xi,\sigma $ are one-form fields.
There is a natural projection
\ber
\rho 
: \mathbb {T} \to T~,
\eer
with $\rho (X+\xi)=X$.

If $g$ is a metric on ${\cal M}$, then this gives a metric $g^{-1}$  on $T^* $ and a metric ${\cal H}$ on 
$\mathbb {T}$ with
\ber
{\cal H} ({X+\xi,Y+\sigma})=
g(X,Y)+ g^{-1}(\xi,\sigma)~.
\eer
There is also a natural map
\ber\label{Gmap}
{\cal G}: \mathbb {T} \to \mathbb {T}~,
\eer
with
\ber
{\cal G}^2 =1~,
\eer
defined in terms of the lowering and raising maps $g:T^* \to T $ and $g^{-1}:T \to T^* $
 by
\ber
{\cal G} (X+\xi)= g\xi + g^{-1}X~.
\eer
{ A 2-form $B$ on ${\cal M}$ defines the $B$-{\it transformation} which is a map}
\ber
e^B:\ X+\xi\mapsto
X+\xi+i_XB
\eer
that preserves the metric $\eta$.

The Courant bracket is  defined on smooth
sections of $T\oplus T^*$, and is given by
\ber\label{courant}
\llbracket X+\xi,Y+\eta\rrbracket_C=[X,Y]+{\cal L}_X\eta-{\cal L}_Y\xi-\frac{1}{2}d(i_X\eta-i_Y\xi),
\eer
where $X+\xi,Y+\eta\in C^\infty(T\oplus T^*)$ and ${\cal L}_X$ is the Lie derivative with respect to $X$.
Given a 
3-form $H$, the $H$-twisted Courant bracket $\llbracket ,\rrbracket_H$ is
\ber
\llbracket X+\xi,Y+\eta\rrbracket_H=\llbracket X+\xi,Y+\eta\rrbracket_C+i_Yi_XH.
\eer
If $b$ is a 2-form then  
\ber
\llbracket e^b(W),e^b(Z)\rrbracket_H = e^b\llbracket W,Z\rrbracket_{H+db}\ \ \ \forall\ W,Z\in
C^\infty(T\oplus T^*),
\eer
so that $e^b$ is a symmetry of $\llbracket , \rrbracket_H$ if and only if $db=0$.

The map ${\cal G}$ 
has eigenvalues $\pm 1$ and the $+1$ and $-1$ eigenspaces both have dimension $d$.
It then defines a splitting 
\ber
\mathbb {T}=\mathbb {T}_+\oplus\mathbb {T}_-~,
\eer
of $\mathbb {T}$ into the $\pm 1$ eigenspaces
\ber
\mathbb {T}_\pm:=P_\pm\mathbb {T}~
\eer
 defined by  the projection operators 
\ber
P_\pm :=\half\left(1\pm {\cal G}\right)~.
\eer
The spaces 
\ber
\mathbb {T}_\pm = \{ X+\xi \in C^\infty(T\oplus T^*) : \xi = \pm gX
\}\eer
 have natural identifications with the tangent bundle
\ber\rho_\pm : \mathbb {T}_\pm \to T\eer
given by 
\ber 
\rho_\pm (X\pm gX)=X \, .
\eer

Using the natural matrix notation in which $X+\xi$ is written
\ber\label{element}
 \mathbb {X}=\left(\begin{array}{c}X\\ \xi \end{array}
\right)
\eer
the metrics are represented by the matrices
\ber \eta = \left( \begin{array}{cc}0&1\\
1&0\end{array}\right)
\qquad {\cal H} =\left( \begin{array}{cc}g&0\\
0&g^{-1}\end{array}\right)
\eer
 The map 
${\cal G}$ acts through the matrix
\ber
{\cal G}=\left( \begin{array}{cc}0&g^{-1}\\
g&0\end{array}\right)
\eer
satisfying
\ber
{\cal G} = \eta ^{-1} 
{\cal H}; \qquad 
{\cal G}^2=1
\eer
while the B-map is represented by
\begin{equation}
\exp(B)=\left(\begin{matrix}1&0 \\B&1\end{matrix}\right)~.
\end{equation}
The projection $\rho $ is 
\ber
\rho=
\left(\begin{matrix}1&0 \\0&0\end{matrix}\right)~.
\eer

Given a closed 3-form $H$, in each patch there is a 2-form $b$ such that $H=db$. If $H$ represents an integral cohomology class, $b$ is a gerbe connection.
Then a B-map using $b$ takes a section $W$ of $T\oplus T^*$ to a {local} section
\ber\label{elements}
 \tilde { \mathbb {X}}= e^b  \mathbb {X} = \left(\begin{array}{c}X\\ \xi +bX \end{array}
\right)
\eer
 of a Courant algebroid $E$ with the short exact sequence
\ber
0\to  T^*\to  E\to  T\to  0
\eer
and anchor map $\tilde \rho: E\to T$.
See {\cite{Hitchin:2004ut, Gualtieri:2003dx,Gil1,Cavalcanti:2012fr,bur}} for further discussion.
The $H$-twisted Courant bracket on $T\oplus T^*$ is mapped to the untwisted Courant bracket on $E$:
\ber
\llbracket \tilde { \mathbb {X}},\tilde { \mathbb {Y} }\rrbracket =\llbracket e^b( \mathbb {X}),e^b( \mathbb {Y})\rrbracket  = e^b\llbracket  \mathbb {X}, \mathbb {Y} \rrbracket_H ~.
\eer
The map then gives $\tilde\eta, \tilde{\cal H},\tilde{\cal G}$ on $E$ given by $\tilde\eta= \eta $ and
\ber
\tilde{\cal G}= e^b  {\cal G}e^{-b}, \qquad  \tilde{\cal H}= e^b  {\cal H}e^{-b}
\eer
so that 
\ber
\tilde{\cal G}=\left(\begin{matrix}-g^{-1}b&g^{-1}\\g-bg^{-1}b&bg^{-1}\end{matrix}\right)
=\left(\begin{matrix}1&\\b&1\end{matrix}\right)
\left(\begin{matrix}&g^{-1}\\g&\end{matrix}\right)
\left(\begin{matrix}1&\\-b&1\end{matrix}\right)
\eer
and
\ber
\tilde{\cal H}=\left(\begin{matrix} g-bg^{-1}b & bg^{-1}
\\
-g^{-1}b & g^{-1}  \end{matrix}\right)  = \eta \tilde{\cal G}~.
\eer
$\tilde{\cal H}$ is often referred to as the generalised metric.

The map $\tilde {\cal G}$ 
has eigenvalues $\pm 1$ and  defines a splitting 
\ber
E=E_+\oplus E_-~
\eer
of $E$ into the $\pm 1$ eigenspaces defined by  the projection operators 
\ber
\tilde P_\pm :=\half\left(1\pm \tilde {\cal G}\right)~.
\eer

In discussing generalised complex geometry, we can either work on $E$ with integrability defined with respect to the untwisted Courant bracket {(as in \cite{Hitchin:2004ut,Gualtieri:2003dx})}, or equivalently on $T\oplus T^*$ with  integrability defined with respect to the $H$-twisted Courant bracket. We will adopt the latter strategy. Our results can be transferred to $E$ using the $B$-map.

\section{Generalised Complex Geometry }

A generalised almost complex structure ${\cal J}$ is a bundle  endomorphism of $\mathbb {T}$ which squares to minus the identity and preserves the metric  ${ \eta}$,
\ber\nn\label{jfull}
&&{\cal J}^2=-1\\[1mm]\nn
&&\eta( {\cal J} \mathbb {X}, {\cal J} \mathbb {Y})=\eta( \mathbb {X},\mathbb {Y}).
\eer
The generalised almost complex structure  ${\cal J}$ splits the complexified generalised tangent bundle
into the  $+i$  eigenspace $ {\mathbb {L}}$ and the  $-i$  eigenspace $\overline {\mathbb {L}}$
\ber
\mathbb {T}\otimes {\mathbb C} = {\mathbb {L}} \oplus \overline {\mathbb {L}}~.
\eer
A {\it{ generalised  complex structure}}  {\cite{Hitchin:2004ut}} is a  generalised almost complex structure for which the  subspace
$  {\mathbb {L}} $ is involutive under the $H$-twisted Courant bracket, i.e. one for which
\ber
\llbracket \mathbb {X},\mathbb {Y} \rrbracket_H\in   {\mathbb {L}} ~{\rm  if} ~ \mathbb {X},\mathbb {Y} \in   {\mathbb {L}} ~.
\eer

 A {\it generalised K\" ahler structure}  {\cite{Gualtieri:2003dx}} on a manifold $ {\cal M}$ with metric $g$ can be defined as   a generalised complex structure ${\cal J}_1$
 that commutes with the map ${\cal G}$ defined in \re{Gmap}:
 \ber
 {\cal J} _1{\cal G}={\cal G}{\cal J}_1~.
 \eer
Then
\ber
 {\cal J} _2={\cal G}{\cal J}_1
 \eer
 defines a second generalised complex structure that commutes with
 $ 
 {\cal J} _1$.
 
 Gualtieri \cite{Gualtieri:2003dx} showed that a generalised K\"ahler structure on  $ {\cal M}$
 is equivalent to   a $(2,2)$ or bihermitian geometry   $( {\cal M},g,H,J_\pm)$
 with complex structures $J_\pm$ on $ {\cal M}$. The Gualtieri map gives the generalised complex structures in terms of the 
 complex structures $J_\pm$
 \ber
 \label{gult}
{\cal J}_{1/2}=\frac{1}{2}
\left(\begin{matrix}J_+\pm J_- & -(\omega_+^{-1}\mp\omega_-^{-1}) \\
\omega_+\mp\omega_-&-(J^t_+\pm J^t_-)\end{matrix}\right).
\eer
Here $\omega_\pm $ are the K\"ahler forms $\omega_\pm = g  J_\pm$.

A generalised complex structure ${\cal J}$ on $\mathbb {T}$ gives a generalised complex structure  $\tilde {\cal J}$ on $E$
\ber \tilde {\cal J}= e^b {\cal J} e^{-b}
\eer
such that the $+i $ eigenspace $L$  of  $\tilde {\cal J}$ is involutive with respect to the untwisted Courant bracket, i.e.  for which
\ber
\llbracket \tilde {\mathbb {X}},\tilde {\mathbb {Y}} \rrbracket_C\in L ~{\rm  if} ~ \tilde {\mathbb {X}},\tilde {\mathbb {Y}} \in L~.
\eer
A generalised K\"ahler structure on $E$  
consists of a  $\tilde {\cal J}_1$ commuting with  $\tilde {\cal G}$, and the Gualtieri map is now
\ber
\tilde{\cal J}_{1/2}=\frac{1}{2}\left(\begin{matrix}1&\\b&1\end{matrix}\right)
\left(\begin{matrix}J_+\pm J_- & -(\omega_+^{-1}\mp\omega_-^{-1}) \\
\omega_+\mp\omega_-&-(J^t_+\pm J^t_-)\end{matrix}\right)
\left(\begin{matrix}1&\\-b&1\end{matrix}\right).
\eer

\section{Half Generalised Complex Structures}
\label{HGCS}

The Gualtieri map can be written suggestively as
\ber
{\cal J}_1 = 
P_+ \left(\begin{matrix}J_+ & 0 \\
0 &-J^t_+ \end{matrix}\right)
P_+  
+
 P_- \left(\begin{matrix}J_- & 0 \\
0 &-J^t_- \end{matrix}\right)
P_-  
\eer
so that
\ber
{\cal J}_1={\cal J}_++{\cal J}_-
\eer
where ${\cal J}_\pm$ are endomorphisms of 
 $\mathbb {T}_\pm$:
 \ber
 {\cal J}_\pm : \mathbb {T}_\pm \to \mathbb {T}_\pm
 \eer
 with 
\ber 
{\cal J}_+ =P_+ \left(\begin{matrix}J_+ & 0 \\
0 &-J^t_+ \end{matrix}\right)
P_+ , \qquad
{\cal J}_-=  P_- \left(\begin{matrix}J_- & 0 \\
0 &-J^t_- \end{matrix}\right)
P_- ~. \eer

This motivates defining structures on $\mathbb {T}_\pm$ instead of on $\mathbb {T}$.
We define a {\it {positive chirality half
generalised almost complex structure} }
${\cal J_+}$ as a bundle  endomorphism of $\mathbb {T}_+$ 
 \ber
 {\cal J}_+ : \mathbb {T}_+ \to \mathbb {T}_+
 \eer
 which squares to minus the identity on $\mathbb {T}_+$ and preserves the metric  ${ \eta}$,
 \ber\nn\label{jfull}
&&{\cal J }_+^2=-1_{ \mathbb {T}_+} \\[1mm]\nn
&&\eta( {\cal J_+} \mathbb {X}, {\cal J_+ } \mathbb {Y})=\eta( \mathbb {X},\mathbb {Y}) ~ {\rm for} ~\mathbb {X},\mathbb {Y}\in \mathbb {T}_+~.
\eer
Similarly, we define a negative chirality half
generalised almost complex structure ${\cal J_-}$ as a bundle  endomorphism of $\mathbb {T}_-$ 
which squares to minus the identity on $\mathbb {T}_-$ and preserves the metric  ${ \eta}$.

Using ${\cal J}_+$, we define another two projection operators
\ber\label{jproj}
\Pi_\pm :=\half (1\mp i{\cal J}_+)~.
\eer
This allows a further split
\ber
\mathbb {T}_+ \otimes {\mathbb C} = {\mathbb {L}}_+\oplus \overline {\mathbb {L}}_+
\eer
into the $+i $ eigenspace ${\mathbb {L}}_+ $ and  the $-i $ eigenspace $\overline {\mathbb {L}}_+$.
We define a
positive chirality half
generalised  complex structure ${\cal J_+}$ as a positive chirality half
generalised almost complex structure    for which  
${\mathbb {L}}_+ $ is involutive with respect to the $H$-twisted Courant bracket,  i.e. one for which 
\ber
\llbracket  \mathbb {X}, \mathbb {Y} \rrbracket_H\in {\mathbb {L}}_+ ~{\rm  if} ~ \mathbb {X},\mathbb {Y} \in C^\infty( {\mathbb {L}}_+ )~.
\eer
We will refer to this as the condition that  ${\cal J}_+$ is integrable.

There is a similar construction for negative chirality. A negative
chirality half
generalised almost complex structure ${\cal J}_-$ gives 
a  split
\ber
\mathbb {T}_- \otimes {\mathbb C} = {\mathbb {L}}_-\oplus \overline {\mathbb {L}}_-
\eer
into the $+i $ eigenspace ${\mathbb {L}}_-$ and  the $-i $ eigenspace $ \overline{\mathbb {L}}_-$
and it will be a negative
chirality half
generalised  complex structure if ${\mathbb {L}} _-$ is involutive with respect to the $H$-twisted Courant bracket.

Note that half
generalised  complex structures automatically commute with ${\cal G}$.
A generalised K\"ahler structure corresponds to two half
generalised almost complex structure ${\cal J}_\pm$.
{ The maps $\rho_\pm : \mathbb {T}_\pm \to T$ take the 
half
generalised  complex structures  ${\cal J}_\pm$ on $\mathbb {T}_\pm$ to the complex structures $J_\pm $ on $T$.

There is a similar construction on the Courant algebroid $E$.
A {\it {positive chirality half
generalised almost complex structure} }
$\tilde {\cal J_+}$ is a bundle  endomorphism of $E_+$ 
 \ber
 \tilde {\cal J}_+ : E_+ \to E_+ 
 \eer
 which squares to minus the identity on $E_+ $ and preserves the metric  ${ \eta}$,
 \ber\nn\label{jfull}
&&\tilde {\cal J }_+^2=-1_{ E_+ } \\[1mm]\nn
&&\eta(\tilde {\cal J}_+ W, \tilde{ \cal J} _+ Z)=\eta( W,Z) ~ {\rm for} ~W,Z\in C^\infty(E_+ )~.
\eer
With the decomposition 
\ber
E_+= L_+ \oplus \bar L_+
\eer
where $L$ is the $+i$ eigenspace of $ \tilde {\cal J}_+$,
we define a 
positive chirality half
generalised  complex structure $\tilde {\cal J}_+$ on $E$ as a positive chirality half
generalised almost complex structure  which is integrable, i.e. for which  
$L$ is involutive with respect to the  Courant bracket
\ber
\llbracket  \tilde {\mathbb {X}}, \tilde {\mathbb {Y}} \rrbracket \in L_+~{\rm  if} ~  \tilde {\mathbb {X}}, \tilde {\mathbb {Y}} \in C^\infty( L_+)~.
\eer
There is a similar construction for negative chirality. 

The structures on $E_\pm$ are the $B$-transforms of the structures on $\mathbb {T}_\pm$:
\ber \tilde {\cal J}_\pm= e^b {\cal J}_\pm e^{-b}
\eer
so that explicitly
\ber 
\tilde{\cal J}_+ =e^b P_+ \left(\begin{matrix}J_+ & 0 \\
0 &-J^t_+ \end{matrix}\right)
P_+ e^{-b}, \qquad
\tilde{\cal J}_-=  e^b P_- \left(\begin{matrix}J_- & 0 \\
0 &-J^t_- \end{matrix}\right)
P_-  e^{-b}~.\eer

}

\section{Algebraic Structure}
 
 If $\mathbb {X}\in C^\infty( \mathbb {T}_+)$, then it takes the form
 \ber
 \mathbb {X}
=\left(\begin{array}{c}X\\ gX \end{array}
\right)
\eer
for some vector field $X\in C^\infty( T)$, and the isomorphism $\rho_+ \mathbb {T}_+\to T$
takes $\rho_+:\mathbb {X}\to X$,
 \ber
\rho_+: \mathbb {X}
=\left(\begin{array}{c}X\\ gX \end{array}
\right) \to X ~.
\eer
The identification of $ \mathbb {T}_+$ with $T$ leads to the identification of an automorphism 
$ \cal {A}$ of $ \mathbb {T}_+$ with an automorphism $A$ of $T$. In this section, we will develop some useful formulas making this identification explicit.

An automorphism of $\mathbb {T}$ has the general form
\ber
{\cal A}=\left( \begin{array}{cc}a&c\\
b&d\end{array}\right)
\eer
for some automorphisms
\ber\nn
&a: ~T  \to T  ~,~~~&c: ~T  \to T^*~, \\[1mm]\nn
&b: ~T^* \to T  ~,~~~&d: ~T^*  \to T^*~. \\[1mm]
\eer
This will be an automorphism of $ \mathbb {T}_+$ that leaves $ \mathbb {T}_-$ invariant if
\ber
P_- {\cal A}=0 ,\qquad   {\cal A}P _- =0 
\eer
which implies
$\cal {A}$ must take the form
\ber
{\cal A}=\frac 1 2 \left( \begin{array}{cc}A&Ag^{-1}\\
gA&gAg^{-1}\end{array}\right)~.
\eer
This can be rewritten as
\ber
{\cal A}=P_+ \left( \begin{array}{cc}A&0\\
0&gAg^{-1}\end{array}\right)~.
P_+
\eer

Then 
\ber
{\cal A}
\left(\begin{array}{c}X\\ gX \end{array}
\right)= \left(\begin{array}{c}AX\\ gAX \end{array}
\right)
\eer
and the automorphism ${\cal A}: \mathbb {X} \to {\cal A} \mathbb {X} $
corresponds 
to the automorphism
\ber A: ~T  \to T 
\eer
of $T$ taking $A:X\to AX$.

For
$\mathbb {X},\mathbb {Y}\in C^\infty(\mathbb {T}_+)$ with
 \ber
 \mathbb {X}
=\left(\begin{array}{c}X\\ gX \end{array}
\right), \qquad
 \mathbb {Y}
=\left(\begin{array}{c}Y\\ gY \end{array}
\right)~,
\eer
we have
\ber
\eta (\mathbb {X},\mathbb {Y}) =  {\cal H} (\mathbb {X},\mathbb {Y}) = 2 g(X,Y)~.
\eer
Then
${\cal A}$ is orthogonal
\ber
\eta ({\cal A}\mathbb {X},{\cal A}\mathbb {Y}) =  \eta (\mathbb {X},\mathbb {Y})
\eer
if and only if $A$ satisfies the orthogonality condition
\ber
g(AX,AY)=g(X,Y)~.
\eer
This is equivalent to
\ber
gAg^{-1}= (A^t)^{-1}
\eer
so that   an orthogonal transformation ${\cal A}$ takes the form
\ber
{\cal A}=P_+ \left( \begin{array}{cc}A&0\\
0& (A^t)^{-1}\end{array}\right)
P_+~.
\eer

A positive chirality half
generalised  almost complex structure ${\cal J_+}$
is then an orthogonal automomorhism of 
 $ \mathbb {T}_+$ satisfying
 ${\cal J }_+^2=-1_{ \mathbb {T}_+}$. It then takes the form
 \ber\label{jplusplus}
{\cal J}_+=P_+ \left( \begin{array}{cc}J_+&0\\
0& (J_+^t)^{-1}\end{array}\right)
P_+
\eer
for an automorphism $J_+:T\to T$
satisfying
\ber
(J_+)^2=-1
\eer
and
\ber
g(J_+X,J_+Y)=g(X,Y)
\eer
so that $J_+$ is an hermitian almost complex structure. As $(J_+)^{-1}=-J_+$, \re{jplusplus} can be rewritten as
\ber\label{jplus}
{\cal J}_+=P_+ \left( \begin{array}{cc}J_+&0\\
0& -J_+^t   \end{array}\right)
P_+~.
\eer
 
Similar formulae apply for $\mathbb {T}_-$.
If
$\mathbb {X}\in
C^\infty(\mathbb {T}_-)$ then
\ber
 \mathbb {X}
=\left(\begin{array}{c}X\\ -gX \end{array}
\right)
\eer
for some $X\in C^\infty(T)$
and an automorphism of $\mathbb {T}_-$
takes the form
\ber
{\cal A}=\frac 1 2 \left( \begin{array}{cc}A&-Ag^{-1}\\
-gA&gAg^{-1}\end{array}\right)
=
P_- \left( \begin{array}{cc}A&0\\
0&gAg^{-1}\end{array}\right)
P_-~.
\eer
For $\mathbb {X},\mathbb {Y}\in C^\infty( \mathbb {T}_-)$, $\eta$ is negative definite
\ber
-\eta (\mathbb {X},\mathbb {Y}) =  {\cal H} (\mathbb {X},\mathbb {Y}) = 2 g(X,Y)
\eer
and ${\cal A}$ is orthogonal if and only $A$ is, as before.
A negative chirality half
generalised  almost complex structure ${\cal J_-}$
is then
\ber
{\cal J}_-=P_- \left( \begin{array}{cc}J_-&0\\
0& -J_-^t   \end{array}\right)
P_-
\eer
for an  hermitian almost complex structure $J_-:T\to T$.

If  $\mathbb {T}$ has a positive chirality half
generalised  almost complex structure ${\cal J_+}$ and  a negative chirality half
generalised  almost complex structure ${\cal J_-}$, then $T$ has hermitian almost complex structures $J_+,J_-$ and 
there are two commuting generalised  almost complex structures
\ber
{\cal J}_1 = 
P_+ \left(\begin{matrix}J_+ & 0 \\
0 &-J^t_+ \end{matrix}\right)
P_+  
+
 P_- \left(\begin{matrix}J_- & 0 \\
0 &-J^t_- \end{matrix}\right)
P_-  
\eer
and
\ber
{\cal J}_2 = 
P_+ \left(\begin{matrix}J_+ & 0 \\
0 &-J^t_+ \end{matrix}\right)
P_+  
-
 P_- \left(\begin{matrix}J_- & 0 \\
0 &-J^t_- \end{matrix}\right)
P_-  
\eer
which can be rewritten as (\ref{gult}).

\section{Integrability}
\label{int}

An almost complex structure $J$ on $T$ splits the tangent bundle into 
$+i$  eigenspace $ \ell $ and the  $-i$  eigenspace $\bar \ell $
\ber
  {T}\otimes {\mathbb C} =\ell \oplus \bar \ell
\eer
and can be used to define the projectors
\ber\label{jproj}
\pi_\pm:=\half (1\mp i{ J})~.
\eer
The almost complex structure $J$ is integrable if $\ell$ is involutive with respect to the Lie bracket,  i.e. if
\ber
[    {X},   {Y} ]\in \ell ~{\rm  if} ~   {X},{Y} \in C^\infty(\ell)~.
\eer
If it is integrable, then it is  a complex structure on $T$. $J$ is integrable if and only if the Nijenhuis tensor defined by
\ber
N(X,Y)= \pi _- ([\pi_+X,\pi_+Y])
\eer
vanishes for all vector fields $X,Y$.

From section \ref{HGCS}, 
a positive chirality half
generalised  almost complex structure ${\cal J_+}$
splits the generalised tangent space
\ber
\mathbb {T}_+ \otimes {\mathbb C} = {\mathbb {L}}_+\oplus \overline {\mathbb {L}}_+
\eer
into 
the $+i $ eigenspace ${\mathbb {L}}_+ $ and  the $-i $ eigenspace $\overline {\mathbb {L}}_+$
and defines the projection operators $\Pi_\mp :=\half (1\pm i{\cal J}_+)$.
It is a
positive chirality half
generalised  complex structure if the subspace ${\mathbb {L}}_+ $  
is involutive with respect to the $H$-twisted Courant bracket, i.e. if
\ber
\llbracket  \mathbb {X}, \mathbb {Y} \rrbracket_H\in {\mathbb {L}}_+ ~{\rm  if} ~ \mathbb {X},\mathbb {Y} \in C^\infty ( {\mathbb {L}}_+ )~.
\eer
This requires the vanishing of the {\it{generalised Nijenhuis tensor}}
\ber\label{genij}
{\cal N} ( \mathbb {X}, \mathbb {Y} )= 
\Pi_- \llbracket  \Pi_+ \mathbb {X}, \Pi_+ \mathbb {Y} \rrbracket_H
\eer
and the vanishing of
\ber\label{Mijenhuis}
{\cal M} ( \mathbb {X}, \mathbb {Y} )= 
P_- \llbracket  \Pi_+ \mathbb {X}, \Pi_+ \mathbb {Y} \rrbracket_H
\eer
which is required for $\llbracket  \Pi_+ \mathbb {X}, \Pi_+ \mathbb {Y} \rrbracket_H$ to be a section of $\mathbb {T}_+$.

For
$\mathbb {X} \in C^\infty( \mathbb {T}_+)$ with
 \ber\label{xinplus}
 \mathbb {X}
=\left(\begin{array}{c}X\\ gX \end{array}
\right)
\eer
it follows from \re{jplus} that
\ber
{\cal J_+} \mathbb {X}
=\left(\begin{array}{c}J_+X\\ gJ_+X \end{array}
\right)
\eer
so that the automorphism $  \mathbb {X} \to {\cal J_+} \mathbb {X}$ of $\mathbb {T}_+$
maps to the automorphism $X\to J_+X$ of $T$.
Then
\ber\label{711}
\Pi_\pm  \mathbb {X}
=\left(\begin{array}{c}\pi_\pm X\\ g\pi_\pm X \end{array}~.
\right)~.
\eer

We have that
\ber
 \mathbb {X}
=\left(\begin{array}{c}X\\ gX \end{array}
\right) \in C^\infty( {\mathbb {L}}_+) ~~~ { \it if ~and~ only~  if} ~ ~~X \in C^\infty ( \ell )~.
\eer

We now turn to the form of the brackets on $ \mathbb {T}_+$.
For
$\mathbb {X},\mathbb {Y}\in C^\infty( \mathbb {T}_+)$ with
 \ber
 \mathbb {X}
=\left(\begin{array}{c}X\\ gX \end{array}
\right), \qquad
 \mathbb {Y}
=\left(\begin{array}{c}Y\\ gY \end{array}
\right)~,
\eer
the Courant bracket takes the simple form
\ber\label{courantt3}
\llbracket  \mathbb {X}, \mathbb {Y} \rrbracket_C
= \left(\begin{array}{c}[X,Y]\\ g[X,Y] \end{array}
\right)
+  \left(\begin{array}{c}0
\\ S(X,Y) \end{array}
\right)
\eer
where $[X,Y]$ is the Lie bracket. 
Here the map   $S: C^\infty(T\otimes T )\to C^\infty(T^*)$
is defined by
\ber
i_Z S(X,Y)=
g( X, \nabla^{(0)} _Z Y)-
g( Y, \nabla^{(0)} _Z X)
\eer
where $\nabla^{(0)} $ is the Levi-Civita connection.
In index notation, 
\ber
S_\mu (X,Y) = X_\nu \nabla^{(0)} _\mu Y^\nu
-Y_\nu \nabla^{(0)} _\mu X^\nu~.
\eer
For the $H$-twisted Courant bracket,
$S$ is replaced in these formulae by
\ber
  S^{(+)}(X,Y) =S(X,Y)+i_Xi_YH
\eer
which has the effect of replacing the connection $ \nabla^{(0)}$ with the connection with torsion
$ \nabla^{(+)}$ given in \re{torcon}.
Then
\ber
i_Z  S^{(+)} (X,Y)=
g( X, \nabla^{(+)} _Z Y)-
g( Y, \nabla^{(+)} _Z X)
\eer
or
\ber\label{smu}
  S^{(+)}_\mu (X,Y) = X_\nu \nabla^{(+)} _\mu Y^\nu
-Y_\nu \nabla^{(+)} _\mu X^\nu~.
\eer
Then 
the H-twisted Courant bracket takes the  form
\ber\label{couranttHH}
\llbracket  \mathbb {X}, \mathbb {Y} \rrbracket_H
= \left(\begin{array}{c}[X,Y]\\ g[X,Y] \end{array}
\right)
+  \left(\begin{array}{c}0
\\ S^{(+)} (X,Y) \end{array}
\right)~.
\eer

We now consider the conditions $\llbracket  \mathbb {X}, \mathbb {Y} \rrbracket_H\in C^\infty( {\mathbb {L}}_+ )$ if $ \mathbb {X},\mathbb {Y} \in C^\infty ({\mathbb {L}}_+ )
$
for  ${\cal J_+}$ to be a 
positive chirality half
generalised  complex structure.
Now from \re{711}, \re{courantt3} and  \re{couranttHH};
\ber\label{couranttH}
\llbracket  \Pi_+ \mathbb {X}, \Pi_+\mathbb {Y} \rrbracket_H
= \left(\begin{array}{c}[\pi_+X, \pi_+Y]\\ g[\pi_+X, \pi_+Y] \end{array}
\right)
+  \left(\begin{array}{c}0
\\ S^{(+)} (\pi_+X, \pi_+Y) \end{array}
\right)~.
\eer
This will be in ${\mathbb {L}}_+ $ if and only if the following two conditions hold
\begin{enumerate}
\item $[\pi_+X, \pi_+Y]\in  C^\infty( \ell )$, i.e. $J_+ $ is a complex structure
\item $S^{(+)} (\pi_+X, \pi_+Y)=0$ for all vector fields $X,Y$.
\end{enumerate}
The second condition is the condition that $\llbracket  \Pi_+ \mathbb {X}, \Pi_+\mathbb {Y} \rrbracket_H \in  C^\infty( \mathbb{T}_+)$
and is equivalent to the vanishing of 
${\cal M} ( \mathbb {X}, \mathbb {Y} )$ in \re{Mijenhuis}, as
\ber
{\cal M} ( \mathbb {X}, \mathbb {Y} )= P_-  \left(\begin{array}{c}0
\\ S^{(+)} (X,Y) \end{array}
\right)~.
\eer
From \re{711} and  \re{smu} 
\ber
  S^{(+)}_\mu (\pi_+ X,\pi_+ Y) =-i( \pi_+ X)_\nu ( \nabla^{(+)} _\mu  J_+^\nu {}_ \rho) Y^\rho
  +
i( \pi_+ Y)_\nu ( \nabla^{(+)} _\mu  J_+^\nu {}_ \rho) X^\rho~.
\eer
 Then the imaginary part of 
 \ber
 S^{(+)}_\mu (\pi_+ X,\pi_+ Y) =0
 \eer
gives
\ber
X^\nu Y^\rho  \nabla^{(+)} _\mu  J_{+ \, \nu \rho} =0~.
\eer
For this to hold for all $X,Y$ requires that $J_+$ is covariantly constant with respect to $\nabla^{(+)} $
\ber
\nabla^{(+)} _\mu  J_{+ \, \nu \rho} =0~.
\eer
If this holds, then it follows from \re{genij} and \re{couranttH} that
 the generalised Nijenhuis tensor is given in terms of the Nijenhuis tensor
\ber
{\cal N} ( \mathbb {X}, \mathbb {Y} )= \left(\begin{array}{c}N(X,Y) \\ gN(X,Y) \end{array}
\right)
\eer
and this will vanish if and only if $N(X,Y)=0$.

We then have the result that 
a positive chirality half
generalised    complex structure ${\cal J_+}$ is equivalent to a hermitian complex structure $J_+$ that is covariantly constant with respect to $\nabla^{(+)} $, $\nabla^{(+)}  J_{+  } =0$.
Similar arguments lead to the result that 
a negative chirality half
generalised    complex structure ${\cal J_-}$ is equivalent to a hermitian complex structure $J_-$ that is covariantly constant with respect to $\nabla^{(-)} $, $\nabla^{(-)}  J_{- } =0$.
To see this, note that
for
$\mathbb {X},\mathbb {Y}\in  C^\infty( \mathbb {T}_-)$ with
 \ber
 \mathbb {X}
=\left(\begin{array}{c}X\\ -gX \end{array}
\right), \qquad
 \mathbb {Y}
=\left(\begin{array}{c}Y\\- gY \end{array}
\right)
\eer
the Courant bracket takes the simple form
\ber\label{courantt1}
\llbracket  \mathbb {X}, \mathbb {Y} \rrbracket_C
= \left(\begin{array}{c}[X,Y]\\ - g[X,Y] \end{array}
\right)
-  \left(\begin{array}{c}0
\\ S(X,Y) \end{array}
\right)
\eer
so that
\ber\label{courantt2}
\llbracket  \mathbb {X}, \mathbb {Y} \rrbracket_H
= \left(\begin{array}{c}[X,Y]\\ -g[X,Y] \end{array}
\right)
-  \left(\begin{array}{c}0
\\ S^{(-)} (X,Y) \end{array}
\right)
\eer
where
\ber
  S^{(-)}(X,Y) =S(X,Y)- i_Xi_YH
\eer
so that 
\ber
  S^{(-)}_\mu (X,Y) = X_\nu \nabla^{(-)} _\mu Y^\nu
-Y_\nu \nabla^{(-)} _\mu X^\nu~.
\eer
Then this leads to the condition $\nabla^{(-)}  J_{- } =0$.

\section{Strong K\"ahler with Torsion and  Half Generalised Complex Structures}

We can now give the formulation of Strong K\"ahler with Torsion geometry in terms of generalised complex geometry.
Using the results of the previous sections, we have that a  $(2,1)$ SKT geometry $({\cal M},g,H, J_+)$ is equivalent to  a positive chirality half
generalised  complex structure ${\cal J_+}$  and metric structure ${\cal G}$
on ${\mathbb {T}}$.
The analogue of the Gualtieri map is 
\ber\label{J}
{\cal J}_+ =P_+ \left(\begin{matrix}J_+ & 0 \\
0 &-J^t_+ \end{matrix}\right)
P_+ \eer
which can be written as
\ber\label{finalJ}
{\cal J}_+=
\half\left( \begin{array}{cc}J_+&-(\omega _+)^{-1}\\
 \omega _+&-J_+^t \end{array}\right)
\eer
where $ \omega _+=g J_+$.
Similarly, a  $(1,2)$ SKT geometry $({\cal M},g,H, J_-)$ is precisely a negative chirality half
generalised  complex structure ${\cal J_-}$  and metric structure ${\cal G}$
on ${\mathbb {T}}$, with 
\ber\label{finalJ-}
{\cal J}_-=
\half\left( \begin{array}{cc}J_-&-(\omega _-)^{-1}\\
\omega _-&-J_-^t \end{array}\right)
=
P_- \left(\begin{matrix}J_- & 0 \\
0 &-J^t_- \end{matrix}\right)
P_-~.
\eer

\section{$(p,q)$ Generalised Complex Geometry }

 {From section \ref{herman}, a $(p,q)$ hermitian geometry $({\cal M},g,H , J_+^a, J_-^{a'})$ has $p-1$ complex structures
$J_+^a$ ($a=1,\dots p-1$) and $q-1$ complex structures
$J_-^{a'}$ ($a'=1,\dots q-1$) and for each complex structure $J$, $({\cal M},g,H , J)$ is an SKT space.
Then each complex structure   corresponds to a half
generalised  complex structure, with  $p-1$ half positive chirality
generalised  complex structures ${\cal J}_{+}^a$   on $\mathbb {T}_+ $, and 
  $q-1$ negative chirality half
generalised  complex structures ${\cal J}_-^{a'}$   on  $\mathbb {T}_- $.
This motivates the definition of a  $(p,q)$ generalised complex geometry
$(\mathbb {T}, {\cal G} , H,  {\cal J}_{+}^a, {\cal J}_-^{a'})$ as 
\begin{enumerate}
\item
 The generalised tangent bundle $\mathbb {T}$ with metric 
${\cal G}$ .
\item $\mathbb {T}$ has
 $p-1$  positive chirality half
generalised  complex structures ${\cal J}_{+}^a$
and  $q-1$ negative chirality half
generalised  complex structures ${\cal J}_-^{a'}$ 
 that are each integrable with respect to the $H$-twisted Courant bracket.
 \item
 If $p=4$, the ${\cal J}_{+}^a$ satisfy the quaternion algebra, and if $q=4$ the  ${\cal J}_-^{a'}$ satisfy the quaternion algebra.
\end{enumerate}

The maps $\rho _\pm  :
\mathbb {T} _\pm \to T$ take each half
generalised  complex structure $ {\cal J}_\pm $ on $\mathbb {T} _\pm$ to a complex structure $J_\pm $ on $T$.
This can be made explicit as:
\ber\label{finalJA}\nn
&&{\cal J}_{+}^a=\half\left( \begin{array}{cc}J&-(\omega)^{-1}\\
\omega&-J^t \end{array}\right)_{\!\!\!+}^a\\[2mm]
&&{\cal J}_{-}^{a'}=\half\left( \begin{array}{cc}J&-(\omega)^{-1}\\
\omega&-J^t \end{array}\right)_{\!\!\!- }^{a'}~.
\eer
This then gives us a precise correspondence between a $(p,q)$ generalised complex geometry
$(\mathbb {T}, {\cal G} , H,  {\cal J}_{+}^a, {\cal J}_-^{a'})$ and a $(p,q)$ hermitian geometry $({\cal M},g,H , J_+^a, J_-^{a'})$.

If $p\ge 2$ and $ q\ge 2$, any positive chirality   
half
generalised  complex structure can be combined with any negative chirality one
 to form  a generalised  complex structure, so that we have a basis of $(p-1)(q-1)$  
generalised  complex structures 
\ber {\cal J}^{a b'} = ({\cal J}_{+}^a, {\cal J}_{-}^{b'})
\eer
together with a further set of generalised  complex structures  ${\cal G}  {\cal J}^{a b'} $.
This
gives a generalised K\"ahler structure for $p=q=2$, 3  generalised  complex structures ${\cal J}^{a 1'}$ for $(p,q)=(4,2)$ and 
9  
generalised  complex structures ${\cal J}^{a b'}$ for $(p,q)=(4,4)$.
For $(p,q)=(4,2)$, the space of generalised  complex structures
is a
2-sphere
while for $(p,q)=(4,4)$ it is
$S^2\times S^2$.
The case of 3 generalised complex structures satisfying a quaternion algebra is a generalised hyperk\" ahler structure \cite{bur}.
}

\section{Discussion}

The $(p,q)$ hermitian geometries are characterised by the holonomy groups $Hol(\nabla ^{(\pm)})$ of the connections with torsion $\nabla ^{(\pm)}$.
If the dimension $d$  of the manifold is even, then $p\ge 2$ if $Hol(\nabla ^{(+)})\subseteq U(d/2)$ and 
$q\ge 2$ if $Hol(\nabla ^{(-)})\subseteq U(d/2)$, while if $d$ is a multiple of $4$, then 
$p\ge 4$ if $Hol(\nabla ^{(+)})\subseteq Sp(d/4)$ and 
$q\ge 4$ if $Hol(\nabla ^{(-)})\subseteq Sp(d/4)$. The cases $p>4 $ or $q>4$ are only possible with trivial holonomy.

There are some features of $(p,q)$ hermitian geometries that are particular to certain values of $p$ and $q$. One interesting aspect of $(2,2)$ hermitian geometry, i.e., of generalised K\" ahler geometry, is that there are three Poisson structures on $T$, 
\ber
\sigma_\pm:=(J_+\pm J_-)g^{-1}~,~~~\sigma:=[J_+,J_-]g^{-1}~,
\eer
which can have irregular points that form loci in ${\cal M}$ where the Poisson structures change rank, giving what has been called type change in \cite{Gualtieri:2003dx}. This implies that for $(4,2)$ hermitian geometry there are then 
three sets of three Poisson structures
\ber
\sigma_\pm^a :=(J_+^a \pm J_-)g^{-1}~,~~~\sigma  ^a:=[J_+^a,J_-]g^{-1}~,
\eer
while for $(4,4)$ hermitian geometry there are then 
nine sets of three Poisson structures
\ber
\sigma_\pm^{ab'} :=(J_+^a \pm J_-^{b'})g^{-1}~,~~~\sigma ^{ab'} :=[J_+^a,J_-^{b'}]g^{-1}~,
\eer

Another interesting feature of generalised K\" ahler geometry is that it has a generalised potential $K$ that determines all geometric quantities. For commuting complex structures, 
\ber
[J_+,J_-]=0~,
\eer
the expressions for the metric and $b$-field are linear in the 2nd derivatives of the scalar function $K$ \cite{Gates:1984nk}, while they are non linear when the complex structures  do not commute \cite{Lindstrom:2007qf, Lindstrom:2007xv, Bredthauer:2006hf}. A proof of this relies on the superspace formulation of the sigma model   away from irregular points. An alternative proof of the existence of $K$ for a class of GKGs that  {\em does} include irregular points was recently given in \cite{Bischoff:2018kzk}. 
For  $(p,q)$ models with $p,q\geq 2$, the scalar potential  governing the geometry is further constrained  to allow for enhanced supersymmetry. In certain classes of models for  which   the supersymmetry algebra closes without use of equations of motion (i.e. off-shell), the general model can be found explicitly, giving  a  general  construction of the corresponding  potential and hence the local geometry \cite{Gates:1984nk}, \cite{Hull:2016khc, Hull:2017hfa}.

For  SKT geometry  there is only one  complex structure
 and so no Poisson structures or irregular points. The potential that determines the geometry again follows from the sigma model and is a complex vector potential $(k_\alpha, k_{\bar\alpha})$ \cite{Hull:1985jv}, \cite{Hull:1985zy}. For recent discussions of such $(2,0)$ and $(2,1)$ models, see \cite{Hull:2016khc, Hull:2017hfa}.
For $(4,1)$ hermitian geometry  (hyperk\" ahler with torsion) the vector potential is further constrained, and has been studied in \cite{Hull:2016khc, Hull:2017hfa}.

\vspace{2cm}

\noindent{\bf Acknowlegement}: \\

\noindent We thank Rikard von Unge for his involvement during the early stages of this collaboration, and we thank
    Gil Cavalcanti for comments.
U.L. gratefully acknowledges the hospitality of the theory group at Imperial College, London. We both acknowledge the stimulating atmosphere at the ``Corfu EISA workshop on Dualities and Generalized Geometry, Sep. 2018.
 This work was supported by the EPSRC programme grant ``New
Geometric Structures from String Theory'',  EP/K034456/1.


\begin{thebibliography}{99}

\bibitem{Gates:1984nk}
  S.~J.~Gates, Jr., C.~M.~Hull and M.~Ro\v cek,
  ``Twisted Multiplets and New Supersymmetric Nonlinear Sigma Models,''
  Nucl.\ Phys.\ B {\bf 248} (1984) 157.
  
  \bibitem{Hitchin:2004ut}
N.~Hitchin,``Generalized Calabi-Yau manifolds'',
Quart.\ J.\ Math.\ Oxford Ser.\ {\bf 54} (2003) 281
math/0209099 [math-dg].

\bibitem{Gualtieri:2003dx}
M.~Gualtieri, ``Generalized complex geometry'',
math/0401221 [math-dg].

\bibitem{Hull:1985jv} 
  C.~M.~Hull and E.~Witten,
``Supersymmetric Sigma Models and the Heterotic String,''
  Phys.\ Lett.\ B {\bf 160}, 398 (1985).

\bibitem{Hull:1985zy}
  C.~M.~Hull,
  ``$\sigma$ Model Beta Functions and String Compactifications,''
  Nucl.\ Phys.\ B {\bf 267} (1986) 266.

 \bibitem{Gil1}
 G.~R.~Cavalcanti,
  ``Reduction of metric structures on Courant algebroids''
  arXiv:1203.0497v1 [math.DG], 2012.
  
\bibitem{Cavalcanti:2012fr}
  G.~R.~Cavalcanti,
  ``Hodge theory and deformations of SKT manifolds,''
  arXiv:1203.0493 [math.DG].

\bibitem{Hull:1986hn}
  C.~M.~Hull,
  ``Lectures On Nonlinear Sigma Models And Strings,''
  Lectures given at  the
Vancouver  Advanced Research Workshop, published in  {\it Super  Field
Theories}    (Plenum,  New  York,  1988),  edited  by  H.~Lee   and
G.~Kunstatter.

 \bibitem{bur}
  H.~Bursztyn, G.~R.~Cavalcanti, M.~Gualtieri,
  "Generalized K\"ahler and hyper-K\"ahler quotients"
  Poisson geometry in mathematics and physics, 61--77, Contemp. Math., 450, Amer. Math. Soc., Providence, RI, 2008
  	arXiv:math/0702104.
	
\bibitem{Howe:1996kj} 
  P.~S.~Howe and G.~Papadopoulos,
  ``Twistor spaces for HKT manifolds,''
  Phys.\ Lett.\ B {\bf 379}, 80 (1996)~,
  [hep-th/9602108].


\bibitem{Buscher:1987uw}
  T.~Buscher, U.~Lindstr\"om and M.~Ro\v cek,
  ``New Supersymmetric $\sigma$ Models With {Wess-Zumino} Terms,''
  Phys.\ Lett.\ B {\bf 202} (1988) 94.





\bibitem{Lindstrom:2005zr}
  U.~Lindstr\"om, M.~Ro\v cek, R.~von Unge and M.~Zabzine,
  ``Generalized K\"ahler manifolds and off-shell supersymmetry,''
  Commun.\ Math.\ Phys.\  {\bf 269} (2007) 833
  [hep-th/0512164].




  




  
 
  
   
 	
\bibitem{Lindstrom:2007qf}
  U.~Lindstr\"om, M.~Ro\v cek, R.~von Unge and M.~Zabzine,
  ``Linearizing Generalized K\"ahler Geometry,''
  JHEP {\bf 0704} (2007) 061~,
  [hep-th/0702126].
  
\bibitem{Lindstrom:2007xv}
  U.~Lindstr\"om, M.~Ro\v cek, R.~von Unge and M.~Zabzine,
  ``A potential for Generalized K\"ahler Geometry,''
  IRMA Lect.\ Math.\ Theor.\ Phys.\  {\bf 16} (2010) 263~,
  [hep-th/0703111].
  
\bibitem{Bredthauer:2006hf}
  A.~Bredthauer, U.~Lindstr\"om, J.~Persson and M.~Zabzine,
  ``Generalized K\"ahler geometry from supersymmetric sigma models,''
  Lett.\ Math.\ Phys.\  {\bf 77} (2006) 291~,
  [hep-th/0603130].
  
\bibitem{Bischoff:2018kzk}
  F.~Bischoff, M.~Gualtieri and M.~Zabzine,
  ``Morita equivalence and the generalized K\"ahler potential,''
  arXiv:1804.05412 [math.DG].

  

  
\bibitem{Hull:2016khc}
  C.~Hull and U.~Lindstr\"om,
  ``All $(4,1)$: Sigma Models with $(4,q)$ Off-Shell Supersymmetry,''
  JHEP {\bf 1703} (2017) 042~,
  [arXiv:1611.09884 [hep-th]].
  
\bibitem{Hull:2017hfa}
  C.~Hull and U.~Lindstr\"om,
  ``All $(4,0)$: Sigma Models with $(4,0)$ Off-Shell Supersymmetry,''
  JHEP {\bf 1708} (2017) 129~,
  [arXiv:1707.01918 [hep-th]].
  
  
  
  

   \end{thebibliography}
\end{document}